# Comparison between Local Ensemble Transform Kalman Filter and PSAS in the NASA finite volume GCM: perfect model experiments

Junjie Liu[1*], Elana Judith Fertig[1], and Hong Li[1]

Eugenia Kalnay[1], Brian R. Hunt[1], Eric J. Kostelich[2], Istvan Szunyogh[1], and Ricardo Todling[3]

[1]University of Maryland, College Park, MD, [2]Arizona State University, [3]NASA-GSFC Global Modeling and Assimilation Office, Greenbelt, MD

Feb. 19[th], 2006

**Abstract**

This paper explores the potential of Local Ensemble Transform Kalman Filter (LETKF) by comparing the performance of LETKF with an operational 3D-Var assimilation system, Physical-Space Statistical Analysis System (PSAS), under a perfect model scenario. The comparison is carried out on the finite volume Global Circulation Model (fvGCM) with 72 grid points zonally, 46 grid points meridionally and 55 vertical levels. With only forty ensemble members, LETKF obtains an analysis and forecasts with lower RMS errors than those from PSAS. The performance of LETKF is further improved, especially over the oceans, by assimilating simulated temperature observations from rawinsondes and conventional surface pressure observations instead of geopotential heights. An initial decrease of the forecast errors in the NH observed in PSAS but not in LETKF suggests that the PSAS analysis is less balanced. The observed advantage of LETKF over PSAS is due to the ability of the forty-member ensemble from LETKF to capture flow-dependent errors and thus create a good estimate of the true background uncertainty. Furthermore, localization makes LETKF highly parallel and efficient, requiring only 5 minutes per analysis in a cluster of 20 PCs with forty ensemble members.

**Introduction**

Three-dimensional variational data assimilation (3D-Var) was adopted for the first time in operational data assimilation at the National Centers for Environmental Prediction (NCEP) with the Spectral Statistical Interpolation (SSI) scheme in 1991 (Parrish and Derber, 1992), and has been proven to be considerably more accurate than the scheme it replaced (Optimal Interpolation, OI). Physical-Space Statistical Analysis System (PSAS), a 3D-Var scheme used operationally at NASA[1], is algebraically equivalent to other 3D-Var schemes, such as the NCEP SSI and the 3D-Var scheme of European Centre for Medium-Range Weather Forecasts (ECMWF) (Courtier et al., 1998), but differs from these schemes in that it is formulated directly in physical space, rather than in model space (Cohn et al., 1998).

In PSAS, like in any other 3D-Var schemes, a constant statistical estimate of the background error covariance is used to represent the background uncertainty. The constant background error covariance makes it difficult to adjust the background (6-hr forecast) to the true state when there are large background errors not well represented by the background error covariance. By contrast, ensemble Kalman filter schemes estimate a flow dependent background error covariance from a time-dependent forecast ensemble. The time changing background error covariance should represent the "errors of the day", if enough ensemble members are used and the model error is small. Furthermore, the ensemble Kalman filter returns an ensemble of analysis states that can be used to initiate the new ensemble forecast. Due to these advantages, as well as its simplicity, the ensemble Kalman filter is considered as candidate for next generation data assimilation system to replace 3D-Var. The ensemble Kalman filter may also be an alternative to 4D-Var schemes, which require the development of the adjoint of the linear tangent model.

---

[1] At the present time, PSAS supports the following operations at NASA: MODIS land team; Aura/MLS; Aura/TES; Aura/HIRDLS; GEOS-Chem group at Harvard University; CERES team; FlashFLUX project team at NASA Langley Research Center; Power project at NASA Langley Research Center; SRB project at NASA Langley Research Center; CALIPSO project. It is also the operational data assimilation system at CPTEC Brazil, and the NWP center in Rome, Italy.

Recently, ensemble Kalman filter (EnKF) schemes have been shown to be able to assimilate real observations effectively. For example, Houtekamer et al. (2005) found the performance of an EnKF scheme to be comparable to the operational 3D-Var scheme when assimilating real observations into the CMC GEM grid model. With more recent changes, the performance became comparable to that of the operational 4D-Var (Houtekamer, 2006, pers. comm.). Whitaker et al. (2004) obtained a better mid-troposphere reanalysis from surface pressure observations with Ensemble Square Root Filter (EnSRF) than with the NCEP SSI on the GFS model at T62L28 resolution. Assimilating all operational observations except radiances, EnKF outperformed the operational SSI (Whitaker et al., 2007 and Szunyogh et al., 2007). These results show promising potential for the ensemble Kalman filter.

Unlike other ensemble Kalman filter schemes that assimilate observations serially (Anderson, 2001, Whitaker and Hamill, 2002, Houtekamer and Mitchell, 2001), the Local Ensemble Transform Kalman Filter (LETKF; Hunt et al., 2007) updates the analysis of each grid point independently by assimilating the observations in the local region centered each grid point simultaneously. The localization approach of LETKF is based on the Local Ensemble Kalman Filter (LEKF) of Ott et al. (2002, 2004). Localization makes the assimilation particularly efficient, as each grid point can be updated in parallel (Szunyogh et al., 2005, 2007). Though adapted from LEKF, the computational cost of LETKF is significantly lower because it solves the analysis equations in the subspace spanned by the ensemble members without using singular value decomposition. This computational efficiency, simplicity of implementation (e.g., it does not require the adjoint of the observational operator and the adjoint of the model dynamics) and its accuracy make LETKF a particularly appealing ensemble Kalman filter scheme.

While assimilation studies based on real observations provide more realistic quantitative estimates of the accuracy of the proposed new schemes (e.g. Whitaker et al., 2007), studies based on simulated observations represent an important step toward better understanding the potential advantages and limitations of the newly proposed schemes (e.g, Szunyogh et al, 2005), since the error statistics are exactly known. In the present study, we compare for the first time an operational 3D-Var system (PSAS) with an ensemble Kalman Filter (LETKF) in a perfect model scenario, by assimilating simulated rawinsondes on the NASA GEOS4 finite volume General Circulation Model (fvGCM) of Lin (2004). We examine and explain the differences in the performance of these two schemes and address the question of the ensemble size required to obtain an accurate background error covariance in the LETKF.

The paper is organized as follows: The LETKF and PSAS assimilation schemes are briefly described in Section 2. The fvGCM is described in Section 3. The experimental setup is explained in Section 4. Results comparing PSAS and LETKF are shown in Section 5. Section 6 is a brief discussion of the origin of differences in the performance of LETKF and PSAS. Section 7 discusses the number of ensemble members required to obtain an accurate estimate of the error covariance in the LETKF scheme. Section 8 is a summary and discussion.

## 2. Assimilation schemes

### 2.1 Physical-Space Statistical Analysis System (PSAS)

PSAS (Cohn et al., 1998) solves the standard analysis equations to minimize the analysis error variance:

$$\mathbf{x}^a = \mathbf{x}^b + \mathbf{K}[\mathbf{y}^o - h(\mathbf{x}^b)] \tag{1}$$

$$\mathbf{K} = \mathbf{P}^b \mathbf{H}^T (\mathbf{H} \mathbf{P}^b \mathbf{H}^T + \mathbf{R})^{-1} \tag{2}$$

Here, $h(\bullet)$ is a nonlinear observation operator transforming the model state variables into observation space, $\mathbf{H}$ is its linearized (Jacobian) operator and $\mathbf{H}^T$ is the transpose (adjoint) of the Jacobian.

Unlike other 3D-Var schemes, PSAS performs most of its computations in the space of observations (Cohn et al., 1998). More specifically, PSAS applies a conjugate gradient (CG) algorithm to solve the system

$$(\mathbf{H} \mathbf{P}^b \mathbf{H}^T + \mathbf{R}) \mathbf{w} = \mathbf{y}^o - h(\mathbf{x}^b) \tag{3}$$

for an intermediate variable $\mathbf{w}$ defined in observation space. This variable is then substituted into Equations 1 and 2 to obtain the updated analysis state

$$\mathbf{x}^a = \mathbf{x}^b + \mathbf{P}^b \mathbf{H}^T \mathbf{w} \tag{4}$$

Together, Equations 3 and 4 are referred to as the PSAS equations. In the specification of the error covariance, only the matrices $\mathbf{HP^bH^T + R}$ and $\mathbf{P^bH^T}$ are calculated and stored. These matrices depend on the observation types.

Because the dimension of the matrices in the above analysis equation depends on the number of observations, the computational cost of PSAS depends primarily on the number of observations, not on the number of model degrees of freedom. It is more efficient to use PSAS than the other 3D-Var systems to assimilate rawinsondes, since there are fewer rawinsonde observations than the number of model degrees of freedom. Because of this efficiency, as well as the availability of the PSAS code to install on our computer system (a 25 dual processor PC cluster with 2.8 GHz dual Xeon speed), we chose PSAS as the standard 3D-Var assimilation scheme with which to compare LETKF. Also, the version of PSAS available for this study was developed to assimilate geopotential height observations, so we assimilate geopotential height observations in comparing PSAS with LETKF.

**2.2 Local Ensemble Transform Kalman Filter**

A detailed description of LETKF is given in Hunt et al. (2007). In the following sub-sections, we briefly summarize the algorithm and its application on the fvGCM. For this application, we first determine the forecast and observation state (Section 2.2.1), then do localization around each grid point (Section 2.2.2), next use local information to update the central grid point of each local region in parallel (Section 2.2.3), and finally combine the analysis at every grid point to obtain a global analysis for each ensemble member (Section 2.2.4).

**2.2.1 Global ensemble forecasts**

First, an ensemble of *k* forecasts, the *i*th of which is denoted by $\mathbf{x}_g^{b(i)}$, is created by integrating the fvGCM from each analysis ensemble member valid at the previous analysis time. Then, each of the forecasts is transformed into observation space by applying the observation operator. The output is denoted as $\mathbf{y}_g^{b(i)} = h(\mathbf{x}_g^{b(i)})$, where the sub-index $g$ represents global vectors.

**2.2.2 Localization and parallelization**

A distinguishing characteristic of LETKF is its localization scheme. Most ensemble Kalman filter techniques introduce covariance localization (Houtekamer and Mitchell, 2001; Hamill et al., 2001) to avoid the spurious long-distance correlations introduced by sampling with a limited number of ensemble members. The version of LETKF used in this paper addresses this problem by cutting off a local region around each grid point (Ott et. al., 2002, 2004; Hunt et al., 2006), such as the local box shown in the Figure 1a. The analysis is performed in the local box, and only the information within this local box is used to update the center point[2]. There is substantial overlap between different local regions corresponding to neighboring grid points, such as the two local boxes centered at the black and grey dots in Figure 1a. The overlap between adjacent local boxes ensures spatial continuity of the analysis.

Because the state is updated independently at each grid point, the cost of LETKF can be dramatically reduced by parallel computation. For our application, the parallel implementation is realized by separating the whole globe into a number of latitude strips based on the number of available processors. The analysis of each latitude strip is computed independently on different processors.

---

[2] Alternatively, the localization can be based on the choice of the observations used at each grid point (Hunt et al. 2007). This approach has some advantages over the box localization adopted here, especially near the poles.

Besides its advantages for parallel implementation, performing such grid point localization greatly reduces the cost of the assimilation. All the vectors presented in the next subsection are reduced from global to local size. In this way, rather than having to assimilate observations serially (one after the other) as in several other ensemble Kalman filter techniques (Tippett et al., 2003), LETKF assimilates all relevant observations simultaneously (Ott et al., 2002, 2004; Szunyogh et al., 2005). Simultaneous assimilation, which allows for observation error correlations in space, is particularly important when the observation coverage is dense and correlated, such as for satellite observations. It can also assimilate observations at the appropriate time when the 4D-LETKF extension is used, which allows for observation error correlations in time as well (Hunt et al., 2004; Hunt et al., 2007; Kalnay et al., 2007).

### 2.2.3 Local analysis

Within each latitude strip, LETKF is performed for all the local boxes around each grid point in the latitude strip sequentially. As described by Szunyogh et al. (2005), at each grid point, the local background vector $\mathbf{x}^{b(i)}$, the corresponding local background vector in observation space $\mathbf{y}^{b(i)}$, and the local observation vector $\mathbf{y}^{o}$ only include the variables within the local box. Different localizations may be chosen for different observations, such as satellite radiances (Fertig et al., 2007). Within the local box, the background state is defined as the ensemble mean of the local forecast vectors:

$$\bar{\mathbf{x}}^b = k^{-1}\sum_{i=1}^{i=k}\mathbf{x}^{b(i)} \tag{5}$$

Unlike LEKF (Ott et. al., 2002, 2004; Szunyogh et al., 2005), LETKF does not calculate the background error covariance explicitly. Following Hunt et al. (2007), the analysis perturbation vector in this local box is given by

$$\mathbf{X}^a = \mathbf{X}^b(\tilde{\mathbf{P}}^a)^{\frac{1}{2}} \tag{6}$$

where $\tilde{\mathbf{P}}^a$, the analysis error covariance in the ensemble space, is given by

$$\tilde{\mathbf{P}}^a = \left[(k-1)\mathbf{I} + \mathbf{Y}^{bT}\mathbf{R}^{-1}\mathbf{Y}^b\right]^{-1} \tag{7}$$

Here, $\mathbf{X}^b$ is the matrix of background ensemble perturbations in local model space, that is, its *i*th column is given by $\mathbf{X}^{b(i)} = \mathbf{x}^{b(i)} - \bar{\mathbf{x}}^b$, $\mathbf{X}^a$ is the corresponding matrix of analysis ensemble perturbations and $\mathbf{Y}^b$ the matrix of ensemble background perturbations in observation space with the *i*th column given by $\mathbf{Y}^{b(i)} = \mathbf{y}^{b(i)} - \bar{\mathbf{y}}^b$, with $\bar{\mathbf{y}}^b = k^{-1}\sum_{i=1}^{i=k}\mathbf{y}^{b(i)}$. Because $\mathbf{Y}^b$ is formulated using the nonlinear observation operator $h(\mathbf{x}^{b(i)}) - h(\bar{\mathbf{x}}^b) \approx \mathbf{H}(\mathbf{x}^{b(i)} - \bar{\mathbf{x}}^b)$, LETKF does not require either the Jacobian $\mathbf{H}$ of the observation operator or its adjoint $\mathbf{H}^T$, unlike 3D-Var or 4D-Var methods. The ensemble mean state of LETKF is updated by the equation:

$$\bar{\mathbf{x}}^a = \bar{\mathbf{x}}^b + \mathbf{X}^b\tilde{\mathbf{P}}^a\mathbf{Y}^{bT}\mathbf{R}^{-1}(\mathbf{y}^o - \bar{\mathbf{y}}^b) \tag{8}$$

The analysis ensemble is given by adding the analysis mean to the analysis perturbations: $\mathbf{x}^{a(i)} = \mathbf{X}^{a(i)} + \bar{\mathbf{x}}^a$.

### 2.2.4 Global analysis ensemble

The local analysis described above returns the analysis ensemble for the center grid point of the local box. The analyses for each grid point in the latitude strip are then connected to return a single file for each strip. The global analysis for each ensemble member is extracted by combining those files. These global analysis states are then used to initiate the ensemble forecast discussed in Section (2.2.1).

To summarize, we plot a schematic flow chart of LETKF scheme (Fig. 1). The forecast model and the observation operator are applied globally, once for each ensemble member. The output of the ensemble forecasts and the observation operator, together with the observations are the input to the LETKF scheme. The analysis ensemble, which is the output from LETKF, is used as initial condition for the next fvGCM ensemble forecast, and the cycle continues.

## 3. NASA fvGCM

The dynamical core of the NASA GEOS4 model is the finite volume General Circulation Model (fvGCM), an atmospheric model developed by Lin (2004) with highly accurate numerical discretization. The fvGCM solves the governing equations by employing a Lagrangian vertical coordinate. Unlike many models that forecast surface pressure, the NASA fvGCM forecasts the pressure thickness ($\delta p$) between vertical model levels and updates surface pressure (*Ps*) as a diagnostic variable. The fvGCM also forecasts zonal wind (u), meridional wind (v), scaled potential temperature (θ), and specific humidity (q).

The version of the fvGCM employed in our experiments (GEOS4) has a horizontal resolution of 5° longitude and 4° latitude (72 zonal and 46 meridional grid points). The model has 55 vertical levels and includes a very high top at 0.01hPa. We note that the horizontal resolution is coarser than that used operationally, but this allows performing many needed experiments under our limited computational resources.

## 4. Simulated observations and experimental design

The assimilation experiments described in this study were conducted in the perfect model scenario. A nature run, representing the true state of the atmosphere, was created by running the NASA fvGCM for three months from the operational analysis of December 16, 2002. Simulated rawinsonde observations were obtained by converting the true model state to rawinsonde variable types, interpolating this converted true state to the real rawinsonde locations, and then adding zero-mean Gaussian distributed noise with standard deviations similar to the operationally assumed rawinsonde errors (Table 1).

To have a direct comparison between LETKF and PSAS, we first created a set of observations including zonal wind, meridional wind and geopotential height. The observation error standard deviations are same as in Table 1. The observations in this first set were at the real rawinsonde observation locations shown in Figure 2a for 00Z. A similar number of observations were available at 12Z, but there were far fewer observations available at 06Z and 18Z (not shown). In creating geopotential height observations, we ignored the vertical error correlations present in reality.

Currently, geopotential height is no longer assimilated in operational centers, such as NCEP and ECMWF. Therefore, an alternative set of observations was created including zonal wind, meridional wind, temperature and surface pressure. This second set of observations is only assimilated by the LETKF scheme since the version of PSAS available in this study was not developed to handle temperature observations. The surface pressure observations were created at the locations of the real surface pressure observation locations, shown in Figure 2b. The standard deviations used for the observational noise of temperature and surface pressure were 1 K and 1 hPa respectively. Based on the hydrostatic balance, temperature and surface pressure observations should provide the same information as the geopotential height observations if they are at the same locations (e.g., Kalnay, 2003). In this case, the second set of observations had more information than the first set of observations because the surface pressure observations were available at greater density than the geopotential height observations.

The initial analysis cycle started at 1800 UTC on 16 December 2002 for PSAS. The initial condition used for this PSAS run was the true state from the nature run at 0000 UTC on 15 January 2003. The LETKF analysis cycle was started on 1800 UTC on 01 January 2003 and used the PSAS analysis as the initial mean state of the ensemble. The initial ensemble members were obtained by adding normally distributed noise to the mean analysis state. The standard deviation of the analysis ensemble perturbations was the same as the standard deviation of the observational noise for observed variables and 0.25K for scaled potential temperature.

PSAS obtains analysis increments of the observed variables, and then converts these increments to update the model state variables (see Equations 3 and 4). LETKF, on the other hand, directly obtains analyses for the model variables. In this study, LETKF directly updated zonal and meridional wind, scaled potential temperature, and surface pressure. Surface pressure is not a prognostic variable, so surface pressure analysis value will not affect the following forecast. Instead, pressure thickness is the related prognostic variable. For simplicity and efficiency in this study, LETKF updated the pressure thickness proportionally to the surface pressure increment for each ensemble member. Specifically, the analysis increment of the pressure thickness at level $k$ for the $i$th ensemble member was given by

$$\frac{\Delta \delta p_k^{a(i)}}{\delta p_k^{b(i)}} = \frac{\Delta Ps_k^{a(i)}}{Ps_k^{b(i)}}, \tag{9}$$

where $\Delta$ indicates the analysis increment of the corresponding variable. Furthermore, neither LETKF nor PSAS updated specific humidity in this study for either LETKF or PSAS to avoid the complexities of assimilating humidity observations (Dee and da Silva, 2003).

To compensate for sampling errors and the effects of nonlinearities in the evolution of the estimation errors, which can lead to an underestimation of the background error covariance and to filter divergence, a multiplicative variance inflation scheme (Anderson and Anderson, 1999) was applied in LETKF. That is, the background error covariance was multiplied by a number larger than 1. In practice, we achieve this as in Hunt et al. (2006) by modifying Equation 7:

$$\tilde{\mathbf{P}}^a = \left[ \frac{(k-1)}{1+\rho} \mathbf{I} + \mathbf{Y}^{bT} \mathbf{R}^{-1} \mathbf{Y}^b \right]^{-1}, \tag{10}$$

which is equivalent to multiplying the background error covariance by a factor of (1+$\rho$). The inflation factor, $\rho$, was tuned to change with level, latitude, and time. At the lower levels the inflation factor was kept constant throughout the assimilation cycle (8% over the entire globe). For the levels above 100hPa, the inflation factor was increased in order to account for model instabilities that appear near the top of the model due to the rigid top boundary condition (Kalnay, 2003, p122). We found experimentally that such larger inflation above 100hPa (increasing linearly from 8% at 100hPa to 100% at the levels above 20hPa) was useful only during the spin-up time. Once the system settles, the inflation was decreased from 8% at 100hPa to about 5% over the polar region. In local boxes where there were no observations, we did not inflate the background, though later studies have found that the analysis improved by inflating the background in these regions (Szunyogh et al., 2007).

The dimensions of the local box were varied spatially to account for inhomogeneous observation coverage and the change of physical distance between grid points with latitudes. The width of the box was increased in the Southern Hemisphere, where observations are sparse. To account for the convergence of the meridians toward the poles, the width of the box was also increased with latitude in both Hemispheres. For example, the horizontal local patch was 7 grid points by 7 grid points in the mid-latitudes in Northern Hemisphere, while it increased to 15 grid points by 7 grid points near the poles. The vertical dimension of the local boxes contained 3 vertical levels, except at the top and the bottom model levels where they contained 1 level.

We tuned the magnitude of this background error covariance to account for the fact that PSAS was originally tuned using real observations and found that the results were not very sensitive to this amplitude. The optimal analysis was obtained when using the operational background error covariance (results not shown).

## 5. Relative performance of LETKF and PSAS

We evaluate the performance of both PSAS and LETKF computing the Root Mean Square (RMS) errors, which are calculated against the true state, for both the analyses and the forecasts. The relative accuracy of the two schemes is examined by comparing the magnitude of the analysis RMS errors, which includes the 500hPa RMS error time series averaged over the globe, the global and time average RMS errors over all vertical levels, and the zonal mean RMS error. We also calculate the percentage improvement of LETKF over PSAS (RMS error difference between LETKF and PSAS normalized by the PSAS RMS error). For the forecasts, we compare the evolution of the RMS error with time in different areas, as well as the representation and impact of the gravity waves on the forecast.

### 5.1 Time series of analysis RMS error

The RMS error time series start from 02 January 2003, when PSAS has already spun-up, and the LETKF analysis cycle begins. As shown in Figure 3a and Figure 3b, after a few days the LETKF analysis (solid line with open circles for the first set of observations and solid line for the second set) has smaller errors than the PSAS analysis (dashed line) for both zonal wind and temperature.

After the initial spin-up period, differences between the RMS errors of each of schemes are significant. The LETKF analysis obtained from assimilating the second set of observations (solid line) has smaller errors than that from assimilating the first set of observations (solid line with open circles). As discussed in Section 4, the second set of observations used here has more information than the first set because there are more surface pressure observations than rawinsonde geopotential height observations. Because LETKF computes realistic error covariance between surface pressure and other variables, the temperature and winds are also significantly improved.

After the spin-up period, the RMS error of the LETKF analysis is not only smaller, but it also shows less variability than that of PSAS. The difference is especially apparent on Feb.12$^{th}$, when PSAS has a large spike in the RMS error. At this time, the RMS error of the LETKF analysis from assimilating the first set of observations (solid line with open circles in Fig. 3) only has slight fluctuations. The LETKF analysis from assimilating the second set of observations is even more stable and accurate (solid line in Fig. 3); it shows no global error spikes after the spin-up time. We will further explore the reasons for this difference in error fluctuations in Section 6.

The RMS error over the Northern Hemisphere (NH, 22ºN–90ºN) (Fig. 4) is much smaller and shows less variability than the global RMS error for both schemes. Because the rawinsonde network is densest in the NH (Fig. 2), the background quickly adjusts to the observations there. Therefore, most of the fluctuations of the errors appear in the regions with low observation density, like the SH (22ºS–90ºS) and oceans.

## 5.2. Vertical and latitudinal structure of the analysis error

Figures 3 and 4 show that the LETKF analysis RMS errors at 500 hPa are smaller than those of PSAS. Figures 5a and 5b show that this holds at all model levels. The RMS error is significantly smaller everywhere than the observational error (Table 1) for both PSAS (dashed line in Figs. 5a and 5b) and LETKF (solid line and solid line with open circles in Fig. 5a and 5b). Assimilating the second set of observations in the LETKF gives much better results than assimilating the first set of observations at all levels. The percentage improvement relative to PSAS is about 40% for zonal wind and 30% for temperature when assimilating geopotential heights (solid line with open circles in Fig. 5c and 5d) and about 60% when assimilating temperature and surface pressure (solid line in Fig. 5c, Fig. 5d). The improvement is larger at the lower levels than at the higher levels.

Figure 6 compares zonally and temporally averaged analysis RMS errors from PSAS and LETKF when both systems assimilate the first observation data set. In the NH, where both schemes are more accurate, LETKF has smaller errors than PSAS. The zonal wind analysis RMS error of LETKF is only between 0.25m/s and 0.5m/s at high latitudes (Fig. 6a), which is about 15% to 25% of the observation error. In most of the Tropics and SH, where the RMS errors are larger, the difference between the RMS error of LETKF and PSAS is also larger (Fig. 6c). Although the RMS error over the Tropics is large for both schemes, the relative improvement of LETKF over PSAS is between 30% and 40% in this region (Fig. 6d). This improvement is important since the tropical regions are the sources of the global energy and hydrological cycle. The percentage improvement is between 40% and 50% through the whole vertical column over the mid-latitudes in the SH (Fig. 6d). However, the percentage improvement becomes smaller over the latitudes beyond 70ºS toward the South Pole, where the LETKF analysis becomes slightly worse than PSAS. The assimilation near the poles was a challenge for the present formulation of LETKF (see footnote on page 8).

### 5.3 Comparison of forecast errors

Since in the perfect model scenario forecast errors originate only from initial errors, and the LETKF analysis has smaller errors than PSAS, better forecasts from LETKF should be expected. This advantage should remain throughout the forecast period, until the errors from both the LETKF and the PSAS forecasts saturate after about two-weeks. It should be noted that if the initial conditions are in balance, growing errors present in the initial conditions should grow exponentially (essentially like bred vectors generated by the analysis cycle, Toth and Kalnay, 1997). If the initial conditions are not in balance, however, we expect that initial errors may not grow, and may even decay, during the geostrophic adjustment period, because of the presence of gravity modes in the analysis. Only after the unbalanced errors disperse, and the growing modes start dominating the error, do we expect to observe exponential growth. Therefore, the error growth observed in different areas of the world provides an indication of the type of errors and the relative balance of the analysis present in the PSAS and LETKF.

As expected, the forecast RMS error of LETKF is smaller than that of PSAS (dashed line) during a five-day forecast period over all the regions (Fig. 7). The forecast from the LETKF analysis assimilating the second set of observations (solid line) has a smaller error than that from PSAS or LETKF assimilating the first set of observations (solid line with open circles). However, different regions show different error growth characteristics. In the NH (Fig. 7a), PSAS errors initially decay, and start growing only after a day, indicating that the PSAS initial conditions are not well balanced. With the same set of observations (the first set of observations), the LETKF starts with smaller errors but they grow faster than those of PSAS, suggesting that when assimilating geopotential heights, our implementation of the LETKF did not completely succeed in suppressing the baroclinic "errors of the day" in the NH. The LETKF analysis using temperatures and surface pressure seems to be the most balanced, and the errors grow more slowly at an approximately constant exponential rate.

In the SH (Fig. 7b), PSAS errors start growing immediately after the analysis, suggesting that the PSAS analysis is more balanced in the SH than in the NH. This is not surprising, since the number of observations in the SH is much smaller than in the NH, and it is the assimilation of observations that cause the PSAS analysis to lose its balance (The assimilation of observations takes place within the subspace spanned by the PSAS background error covariance). The forecast RMS errors of LETKF grow at a similar rate as PSAS.

In the Tropics (22ºS–22ºN, Fig. 7c), the forecast RMS errors of PSAS are almost constant for a couple of days, and then increase linearly. For the LETKF the forecast RMS errors are similar for both data sets. They start smaller than PSAS and grow linearly with time at about the same rate as PSAS. This characteristic linear growth of errors in the Tropics was also observed in Kuhl et al. (2007) in simulations with the NCEP Global Forecasting System (GFS). Unlike extratropical error growth dominated by slow baroclinic waves, tropical errors are dominated by convection, which saturate almost immediately at small scales and slowly propagate to larger scales even in a perfect model scenario (Harlim et al., 2005).

**5.4 Accuracy in representing gravity waves**

These results suggest that we should consider the balance characteristics of both LETKF and PSAS in more detail. Balance is fundamental to the dynamics and predictability of the atmosphere. When in balance, numerical models are in quasi-geostrophic balance in the extra-tropics. However, faster gravity waves can be excited in primitive equation models. Spurious gravity waves are generated by imbalanced initial conditions (e.g., Simmons 1999; Kalnay, 2003). High frequency gravity waves, with high frequency oscillations in the divergence field and surface pressure, are highly dispersive and thus generally not observed with significant amplitudes in the extra-tropics, except for the diurnal and semidiurnal tides.

Variational data assimilation schemes (3D-Var and 4D-Var) maintain the balance in the analysis fields by including geostrophic balance in the background error covariance as well as additional balance constraints in the cost function that they minimize. It is common to apply a balancing algorithm such as nonlinear normal mode initialization (e.g., Daley, 1991), digital filter (Lynch and Huang, 1992) or the Incremental Analysis Update (Bloom et al, 1996) to the analysis in order to eliminate high frequency waves before the next forecast begins, although no such initialization is used in either LETKF or the version of PSAS used in this study.

Ensemble Kalman filter schemes do not need explicit balance constraints or initialization in the grid-point observation data assimilation (Szunyogh et al., 2005). The ensemble analysis minimizes the introduction of spurious gravity waves by computing the analysis as a linear combination of the ensemble forecasts, which are generally well balanced. Fast gravity waves remain in the analysis field only if they are in the background. Although the use of local boxes in the LETKF could lead to imbalances, the large overlap between different local boxes in LETKF is apparently able to minimize the excitement of gravity waves by assimilating similar information in neighboring regions.

To compare the relative ability of both schemes to represent real gravity waves, we plot the true and analyzed horizontal divergence for a period and location (32ºN, 93ºW on 700hPa) where a large amplitude gravity wave in the true dynamical field is observed. The analysis of divergence field from LETKF with the first set of observations (open circles) closely follows the truth (solid line, Fig. 8a), accurately representing the amplitude and phase of the gravity waves that appear in the true field. In agreement with the results of Szunyogh et al. (2005), we do not observe spurious high frequency gravity waves. PSAS (closed circles) also gives a fairly good analysis of the true gravity waves (solid line, Fig. 8b), but the amplitude of the errors is considerably larger than that of LETKF. The gravity wave that appears in the truth has both diurnal and semi-diurnal components, especially around February 14. This structure is apparent in the 2-day forecasts starting on 12Z 14 February plotted every hour (Fig. 8, bottom panel). The forecasted surface pressure shows the diurnal and semidiurnal modes in the truth (crosses), PSAS (full circles), and LETKF (open circles) forecasts. Although both forecasts capture the diurnal and semidiurnal tides, we observe that the initial conditions from the LETKF lead to a more balanced and accurate forecast.

**6. Relationship between analysis increments and background error**

The analysis increments (difference between analysis state and background state) reflect the correction made to the background from the observation information. They are determined by the background error covariance, observation error, and observation innovation (the difference between observation and the background mean state in observation space), as shown in Equation (1) for PSAS and Equation (8) for LETKF. The background error is the difference between background state and the truth, so that the optimal analysis increment should be equal and opposite to the background error.

We analyze the different characteristics and reasons for the difference in the performance of LETKF and PSAS by comparing the analysis increment and background error on 12Z Feb. $12^{th}$, the time at which the largest RMS error difference between LETKF and PSAS occurs (Fig. 3). The largest difference between two schemes is observed over the ocean in the Southern Hemisphere, especially between 30ºS and 80ºS, 150º E and 200º W where there is a deep trough associated with major weather development.

Figure 9 shows the analysis increments (contours) and the background errors (colors) in the region described above. Both schemes extract useful information from the sparse observations, as indicated by the fact that the analysis increments in general have opposite sign to the background errors. However, in LETKF, the analysis increments line up with the background error generally better than in PSAS, even in areas without observations. This is because the background error covariance estimated from the ensemble is able to extrapolate observation information to data sparse regions by accurately reflecting the shape of the errors of the day. Because PSAS has a constant isotropic background error covariance, it cannot estimate abrupt error changes in the shape and amplitude of background error. The structure of the temperature increments in PSAS is significantly different from that of the background error. In PSAS, large analysis increments are observed around the observation locations, not in regions with large background error (Fig. 9b). As a result, PSAS has a worse performance than LETKF.

## 7. The number of ensemble members required in LETKF

As discussed before, the primary advantage of LETKF is that it accounts for the errors of the day through the flow-dependent background error covariance estimated from the background ensemble forecasts. The accuracy of the background error covariance depends on the number of ensemble members. When there are too few ensemble members to capture the background uncertainty, sampling errors may be introduced which would be reflected in an inconsistency between background uncertainty estimated from the ensemble and the actual background error. The accuracy of the background error covariance is not sensitive to the number of ensemble members after enough ensemble members are used to estimate the uncertainty of the dynamics (Ott et al., 2004, Kalnay et al., 2007, Figs. 4 and 5).

We examine the ability of forty ensemble members to adequately represent the true uncertainty by comparing sample ensemble spread to the actual ensemble mean error. Both quantities are time averaged over the second month of the assimilation cycle. The ensemble spread, representing the uncertainty in the ensemble forecast, determines the weight given to both the background and the observations. Its time average is calculated as follows:

$$S = [\frac{1}{T}\sum_{t=1}^{t=T}\frac{1}{k-1}\sum_{i=1}^{i=K}(x_i^b - \bar{x}^b)^2]^{\frac{1}{2}} , \qquad (11)$$

where $x_i^b$ is the i$^{th}$ background ensemble member and $\bar{x}^b$ is the background mean state at one grid point. The error of the ensemble mean is measured by the distance between the ensemble background mean and the true state:

$$V = [\frac{1}{T}\sum_{i=1}^{i=T}(\bar{x}^b - x^t)^2]^{\frac{1}{2}} \qquad (12)$$

where $x^t$ is the true state at one grid point. If the data assimilation is optimal, and there are enough ensemble members to estimate the background error covariance, the background ensemble spread should be same as the error of the ensemble mean.

Figures 10a and 10b show that the 40-member ensemble accurately estimates the shape of the background error. The centers of ensemble spread

(contour) and ensemble mean error (shaded) are approximately at the same locations, and both fields have similar shapes. It is noteworthy that even though both sets of observations provide different amounts of information to LETKF, as discussed before, in each case the ensemble spread is consistent with the ensemble-mean error. We examine the relative amplitude of the ensemble spread and ensemble mean error by calculating their ratio (which ideally should be equal to one). Overall, the 40-member ensemble accurately estimates the magnitudes of background uncertainty over these regions. The ratio of ensemble spread to ensemble-mean error is close to one in data dense regions, such as over land (Fig. 11a and Fig. 11b) and over southern ocean in Figure 11b (the coverage of surface pressure is dense over this region). The spread is slightly larger than the mean error for the first observation set (Fig. 11a), and slightly smaller than the mean error for the second set of observations (Fig. 11b). Since the same inflation factor is applied in the assimilations for both observation data sets, it is apparent that the reduction of the ensemble spread in the second set of observations comes from the denser observational coverage. This suggests that larger inflation factors are required in the data dense region to keep a reasonable ensemble spread. In data sparse regions such as the Tropics, the ratio of ensemble spread to variance is about 1.5-2 (Fig. 11a, Fig. 11b). This larger ratio suggests that the ensemble spread overestimates the background uncertainty, causing the analysis to give more weight to the observations than it should. Further tuning the inflation factor based on the observation coverage may improve the assimilation accuracy, since inflation should be different over data dense and data sparse regions.

Forty ensemble members seem to be enough to adequately capture the background uncertainty under the perfect model scenario. We recognize that the resolution of the model used for this study is much coarser than that of operational models. Accordingly, more than forty ensemble members may be required to estimate the background error covariance for operational models. Nevertheless, Miyoshi et al. (2006, private communication) found that the accuracy of the results do not change much when the ensemble members are increased from forty to eighty with a much higher resolution model of T159L48 model.

## 8. Summary and discussion

In this study we compare the performance of LETKF with an operational 3D-Var scheme, the NASA PSAS analysis system, by assimilating the simulated rawinsonde observations on a finite volume GCM with horizontal resolution of $4^o$ by $5^o$ and 55 levels. With only forty ensemble members, the LETKF analysis shows significantly less RMS error than the PSAS analysis. The relative improvement of the LETKF analysis over that of PSAS is about 30% to 40% in this perfect model scenario. The largest improvement of LETKF over PSAS is found in regions with sparse observations, particularly in the Southern Hemisphere. This result is consistent with Whitaker et al. (2004, 2007) and Szunyogh et al. (2007) finding that ensemble Kalman filters have the most advantage in data sparse regions. The assimilation of simulated temperature and surface pressure observations shows an even more accurate analysis. The 5-day forecast maintains this advantage. The forecast errors starting from the PSAS analysis decrease in the first few hours before they start growing with time, indicating the presence of analysis imbalance that disperses as gravity waves during the initial geostrophic adjustment. By contrast, the initially smaller analysis error in the LETKF analysis grows exponentially, indicating better balance in the initial conditions.

The large improvement of LETKF over PSAS is due to the fact that the background error covariance used in LETKF varies with space and time, which reflects the errors of the day. The constant background error covariance used in PSAS cannot reflect abrupt error changes in the background. As a result, the analysis increments structure are more similar (with opposite sign) to the background errors in LETKF, whereas in PSAS the analysis corrections are more isotropic, and tend to be centered on observation locations.

The LETKF scheme is highly parallel and efficient. The parallel computation characteristic comes from the localization of the LETKF scheme. In the LETKF used here only those observations within a local box are used to update the center grid point. Alternatively, the localization can be based on choosing the observations within a distance to update the center grid point, rather than using a local box (Hunt et al., 2007). The agreement between ensemble spread and ensemble mean error suggests that forty ensemble members used in LETKF are sufficient to capture most of the uncertainty in the global fvGCM forecast. Nevertheless, more ensemble members may be required in a higher resolution model.

Although we used operational 3D-Var analysis system and global model in this study, there are some caveats on the results presented, since they are based on perfect model scenario, in which we have avoided additional challenges associated with the presence of unknown observation and model errors. Also, the observational network only includes rawinsondes, which is much sparser than the operational observation network. Previous research shows that EnKF has more advantage over data sparse region (Whitaker et al., 2004) so that the advantages of LETKF may be smaller with current operational coverage. In addition, the model resolution used is lower than that currently used in operations. Therefore our very encouraging results should perhaps be interpreted as an upper bound for the potential advantage of operational EnKF over 3D-Var analysis.

## Acknowledgements

This work is part of the research towards the fulfillment of the doctorate requirement of the three leading authors, who have contributed equally. This research was partially supported by NSF ATM9328402, NASA NNG04GK29G and NNG04GK78A. We are grateful to Dirceu Herdiez, Arlindo da Silva and Shian-Jiann Lin for very helpful discussions.

LIST OF TABLES

Table 1 Observation error standard deviations varied with vertical levels for simulated zonal wind (U), meridional wind (V) and geopotential height (H) observations. (From NASA PSAS).

| Unit (hPa) | U (m/s) | V (m/s) | H (m) |
|---|---|---|---|
| 1000 | 2.0 | 2.0 | 5.4 |
| 850 | 2.2 | 2.2 | 5.6 |
| 700 | 2.3 | 2.3 | 6.2 |
| 500 | 2.7 | 2.7 | 8.6 |
| 400 | 3.2 | 3.2 | 10.8 |
| 300 | 3.4 | 3.4 | 12.8 |
| 250 | 3.4 | 3.4 | 13.5 |
| 200 | 3.3 | 3.3 | 14.5 |
| 150 | 2.7 | 2.7 | 16.3 |
| 100 | 2.7 | 2.7 | 19.3 |

LIST OF FIGURES

Fig. 1 a. Schematic of the local box used in LETKF. b. Flow chart of the LETKF scheme applied in fvGCM.

Fig. 2 a. Top left panel is the real rawinsonde observation locations (black dots) at 00z. Top right panel is the relative distribution of observational coverage at different pressure levels. The bottom panel is the simulated surface pressure observation location.

Fig. 3 500hPa global average analysis RMS error (y-axis) as function of time (x-axis) for (a) zonal wind and (b) temperature. The dashed line is the result from PSAS, the solid line with open circles is the result of LETKF assimilating the same observations as PSAS (winds and geopotential height), while the solid line is the result from LETKF assimilating wind, temperature and surface pressure observations.

Fig. 4 Same as Fig. 3, except for the analysis RMS error averaged over the Northern Hemisphere (22ºN-90ºN).

Fig. 5 Time mean (averaged over February) of the analysis RMS error averaged over global as function of the vertical levels for zonal wind (a) and temperature (b). The dashed line is the result from PSAS, the solid line with open circles is the result of LETKF assimilating the same observations as PSAS (winds and geopotential height), while the solid line is the result from LETKF assimilating winds, temperature and surface pressure observations. (c) and (d) are relative improvement of LETKF over PSAS for zonal wind and temperature (Solid line is the relative improvment of LETKF assimilating the second observation data set, while solid line with open circles is the relative improvement of LETKF assimilating the first observation data set).

Fig. 6 Zonal average of the time mean zonal wind analysis RMS error from LETKF and PSAS, both assimilating the same winds and geopotential height observations. The contours in (a), (b), and (c) indicate the February average of the true zonal wind field, and the shades indicate the analysis RMS error of: (a) LETKF; (b) PSAS; (c) the difference between LETKF and PSAS analysis RMS errors. (d) is relative improvement of LETKF over PSAS.

Fig. 7 500hPa zonal wind average (averaged over February) forecast RMS error (m/s) as function of the leading time in different regions (a) Northern Hemisphere (22ºN —90ºN); (b) Southern Hemisphere (22ºS—90ºS); (c) Tropics (22ºS—22ºN); (d) global. The dashed line is the result from PSAS, the solid line with open circles is the result of LETKF assimilating the same observations as PSAS (winds and geopotential height), while the solid line is the result from LETKF assimilating the second set observations.

Fig. 8 Top: comparison of the "true" (crosses) and analyzed divergence field every 6 hours at 32ºN, 93ºW on 700hPa (where there is a rawinsonde observation). Left: LETKF (open circles) Right: PSAS (closed circles) assimilating the same observations. The numbers in the text boxes are the RMS differences between the analyses and the truth. The bottom panel is the 2-day surface pressure forecast from 12Z 14 February at 32ºN, 93ºW (crosses show the true pressure, open circles are the LETKF forecast, and full circles are the PSAS forecast). The output interval is every hour.

Fig. 9 500hPa temperature analysis increments (contour) and background error (shaded) for LETKF (left panel) and PSAS (right panel) at 12z Feb 12$^{th}$. The dots represent the rawinsonde observation locations which are geopotential height observations. Both schemes assimilate winds and geopotential height observations.

Fig.10 Average ensemble spread of zonal wind (averaged over February, contour; Unit: m/s) and the ensemble mean error (shades; Unit: m/s) at 500hPa ( (a) is calculated from LETKF with wind and geopotential height observations assimilated, (b) is calculated from LETKF with wind, temperature and surface pressure observations assimilated).

Fig. 11 The ratio of time average ensemble spread and ensemble mean error of zonal wind at 500hPa ((a) is calculated from LETKF with wind and geopotential height observation assimilated, (b) is calculated from LETKF with wind, temperature and surface pressure observations assimilated).

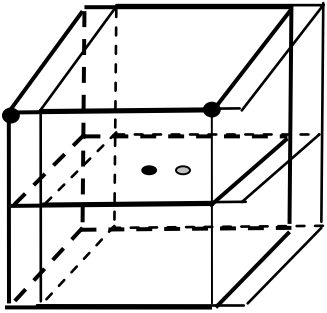

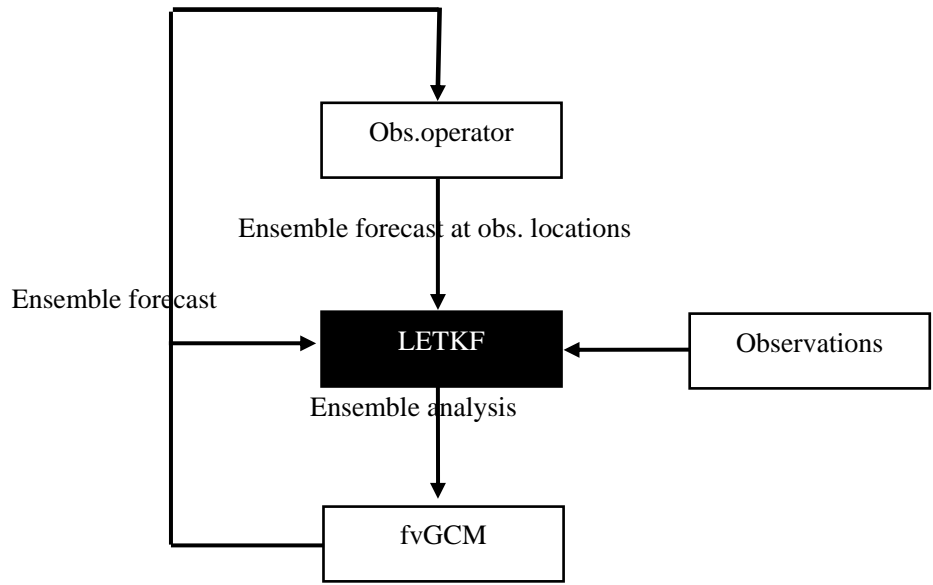

Fig. 1 a. Schematic of the local box used in LETKF. b. Flow chart of the LETKF scheme applied in fvGCM.

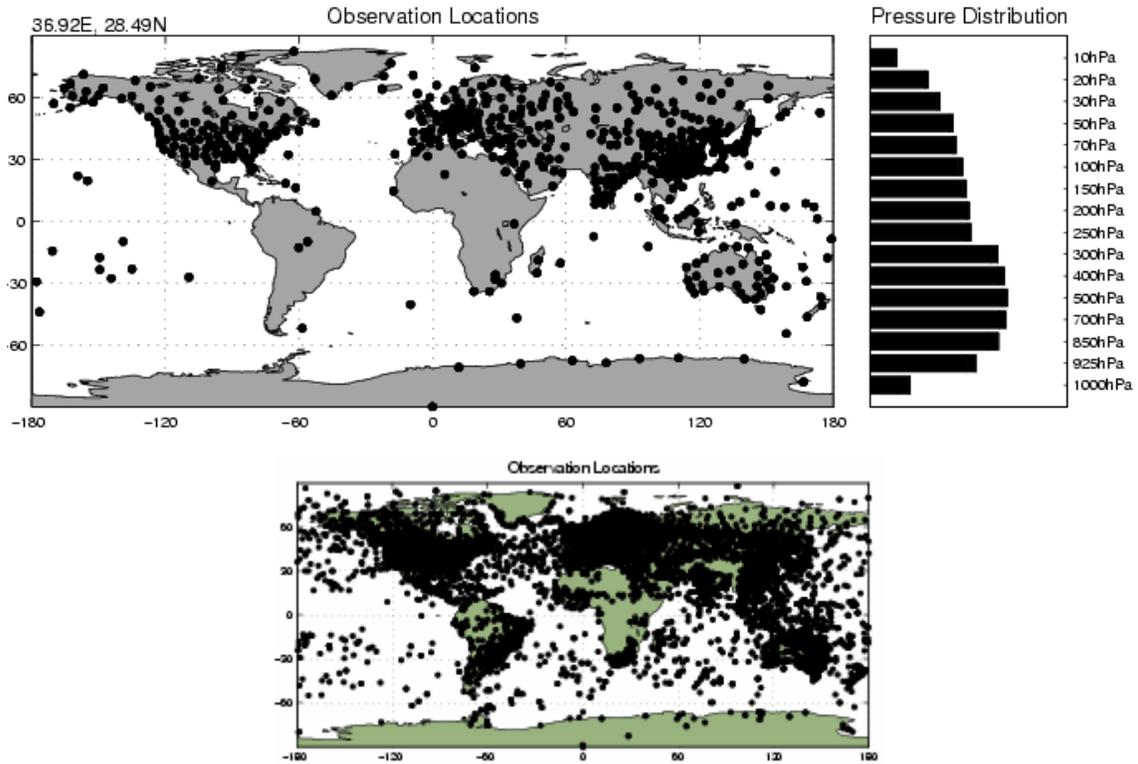

Fig. 2 a. Top left panel is the real rawinsonde observation locations (black dots) at 00z. Top right panel is the relative distribution of observational coverage at different pressure levels. The bottom panel is the simulated surface pressure observation location.

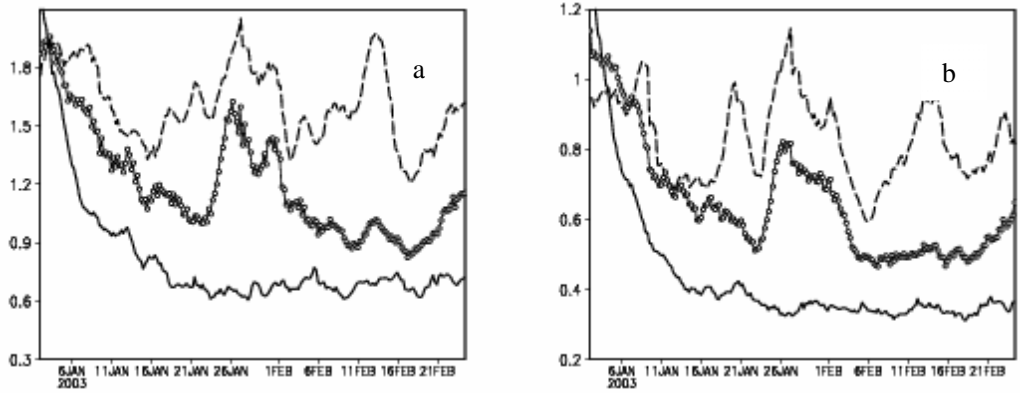

Fig. 3  500hPa global average analysis RMS error (y-axis) as function of time (x-axis) for (a) zonal wind and (b) temperature. The dashed line is the result from PSAS, the solid line with open circles is the result of LETKF assimilating the same observations as PSAS (winds and geopotential height), while the solid line is the result from LETKF assimilating wind, temperature and surface pressure observations.

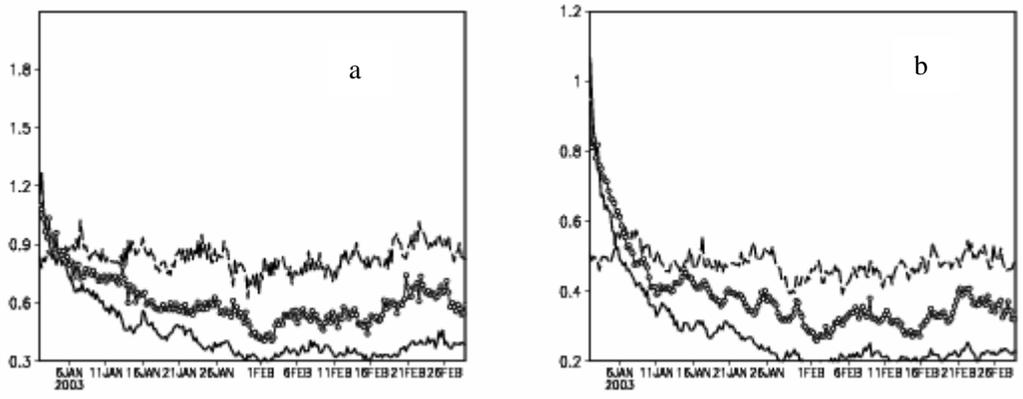

Fig. 4 Same as Fig. 3, except for the analysis RMS error averaged over the Northern Hemisphere (22ºN-90ºN).

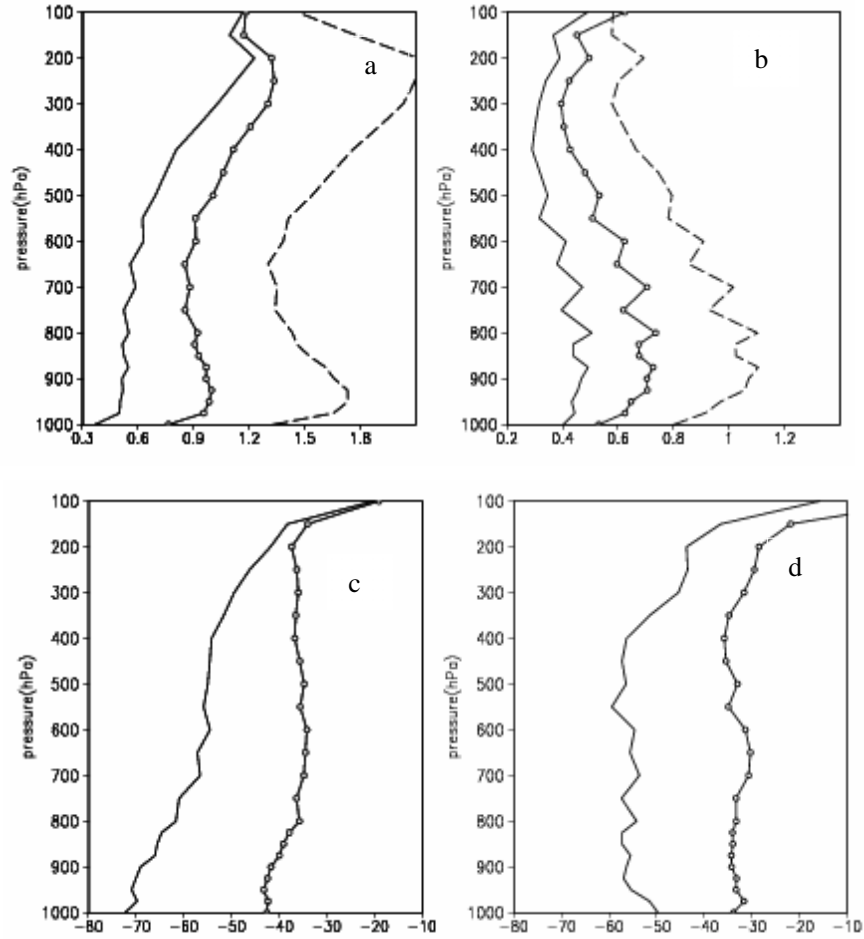

Fig. 5 Time mean (averaged over February) of the analysis RMS error averaged over global as function of the vertical levels for zonal wind (a) and temperature (b). The dashed line is the result from PSAS, the solid line with open circles is the result of LETKF assimilating the same observations as PSAS (winds and geopotential height), while the solid line is the result from LETKF assimilating winds, temperature and surface pressure observations. (c) and (d) are relative improvement of LETKF over PSAS for zonal wind and temperature (Solid line is the relative improvment of LETKF assimilating the second observation data set, while solid line with open circles is the relative improvement of LETKF assimilating the first observation data set).

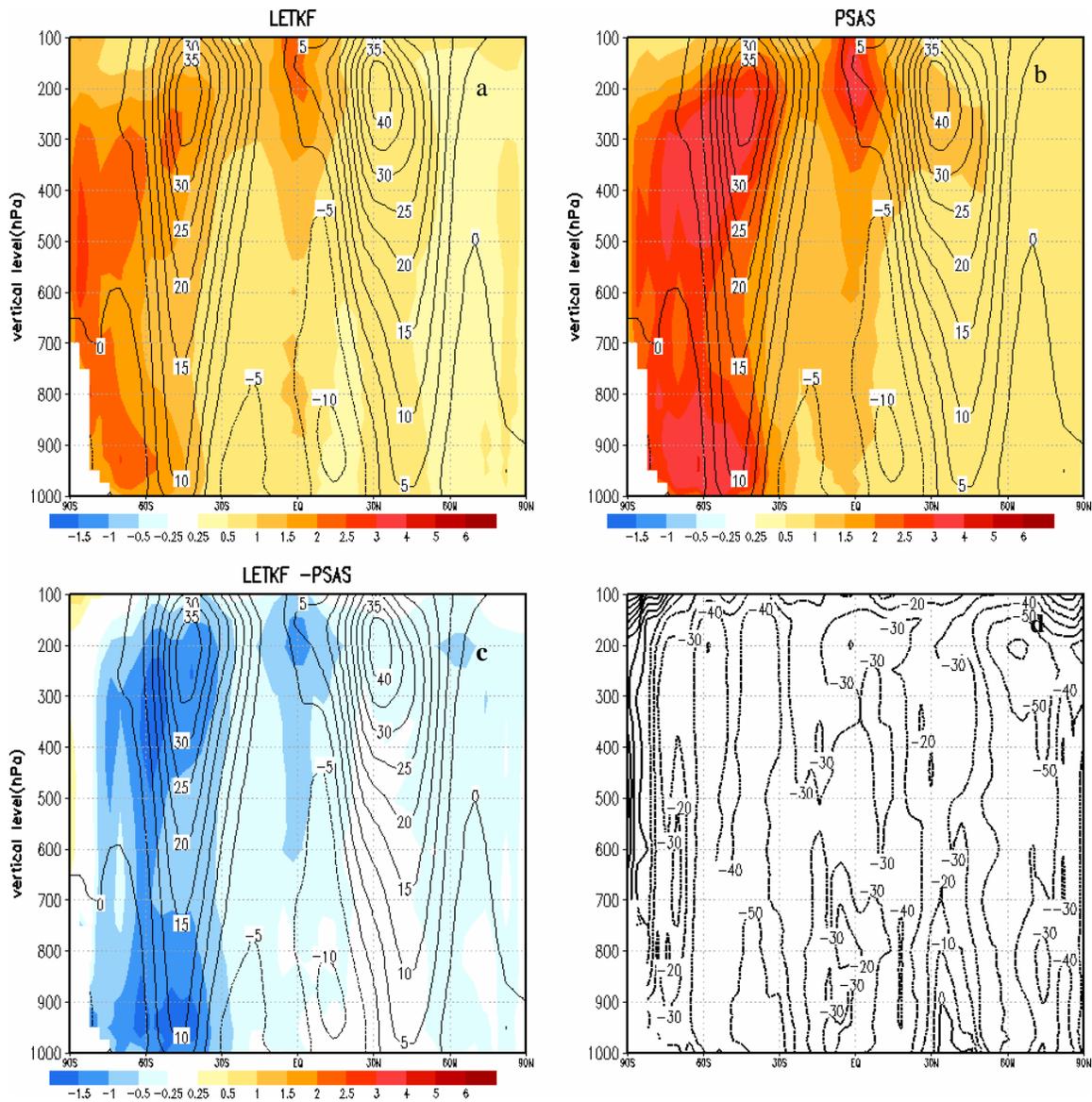

Fig. 6 Zonal average of the time mean zonal wind analysis RMS error from LETKF and PSAS, both assimilating the same winds and geopotential height observations. The contours in (a), (b), and (c) indicate the February average of the true zonal wind field, and the shades indicate the analysis RMS error of: (a) LETKF; (b) PSAS; (c) the difference between LETKF and PSAS analysis RMS errors. (d) is relative improvement of LETKF over PSAS.

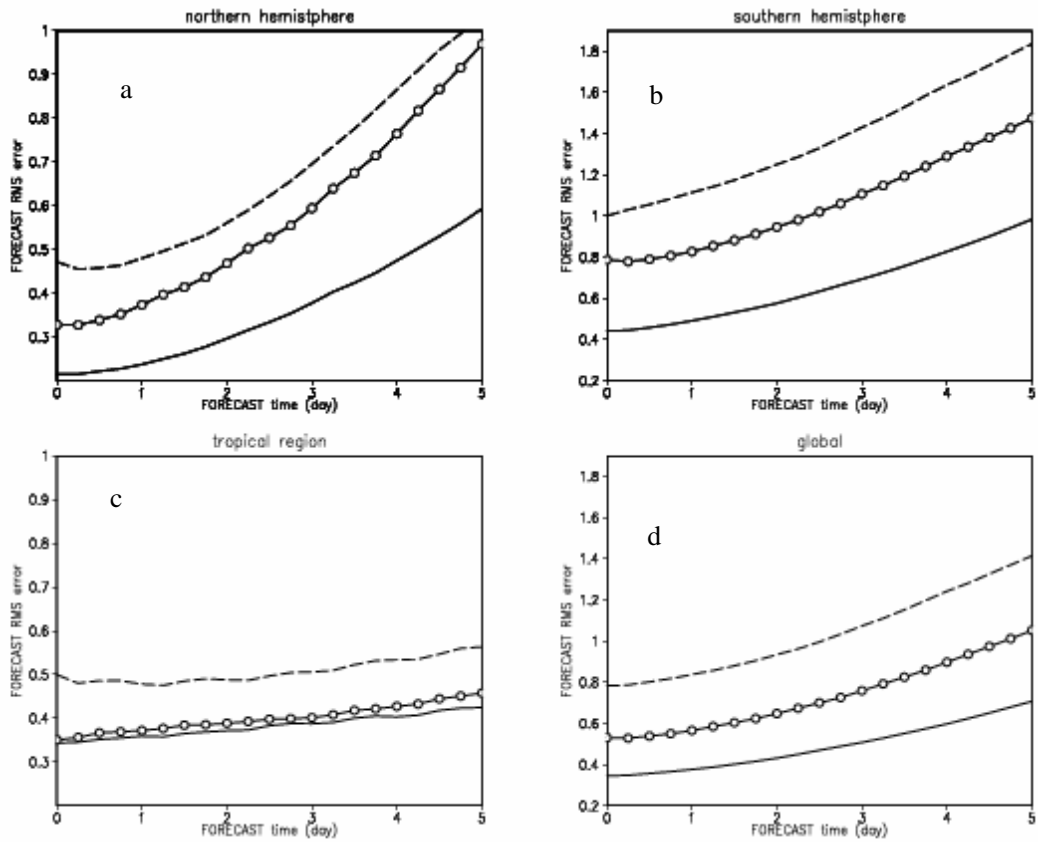

Fig. 7 500hPa zonal wind average (averaged over February) forecast RMS error (m/s) as function of the leading time in different regions (a) Northern Hemisphere (22ºN —90ºN); (b) Southern Hemisphere (22ºS—90ºS); (c) Tropics (22ºS—22ºN); (d) global. The dashed line is the result from PSAS, the solid line with open circles is the result of LETKF assimilating the same observations as PSAS (winds and geopotential height), while the solid line is the result from LETKF assimilating the second set observations.

divergence field from LETKF and truth | divergence field from PSAS and truth

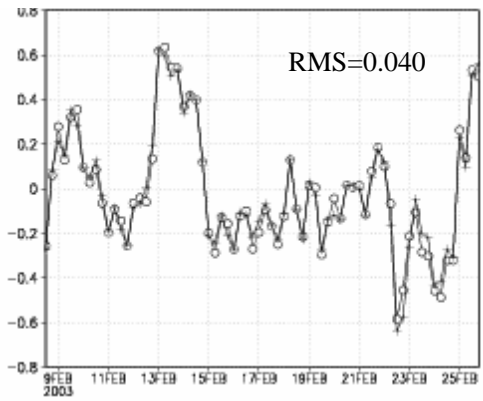
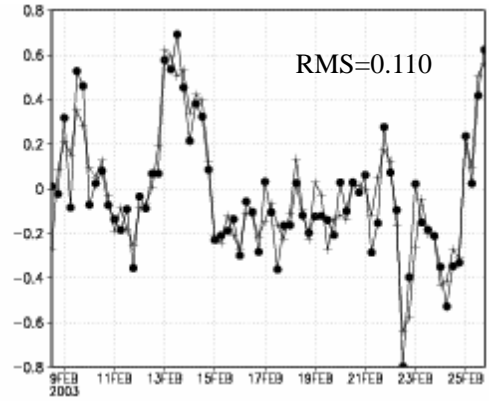
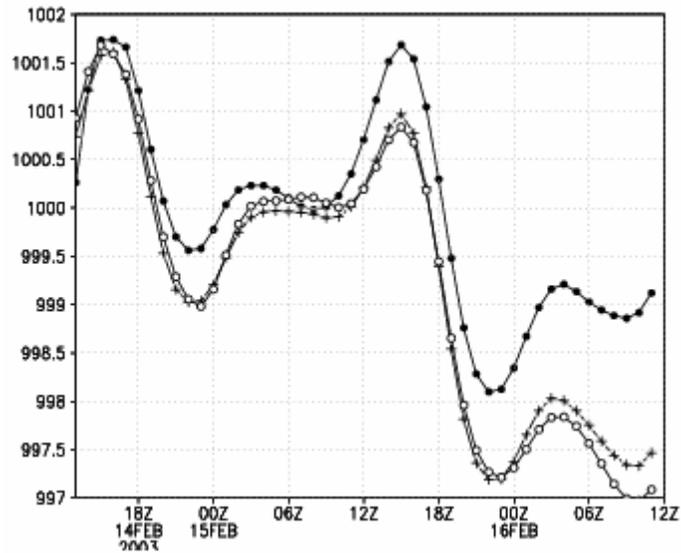

Fig. 8 Top: comparison of the "true" (crosses) and analyzed divergence field every 6 hours at 32ºN, 93ºW on 700hPa (where there is a rawinsonde observation). Left: LETKF (open circles) Right: PSAS (closed circles) assimilating the same observations. The numbers in the text boxes are the RMS differences between the analyses and the truth. The bottom panel is the 2-day surface pressure forecast from 12Z 14 February at 32ºN, 93ºW (crosses show the true pressure, open circles are the LETKF forecast, and full circles are the PSAS forecast). The output interval is every hour.

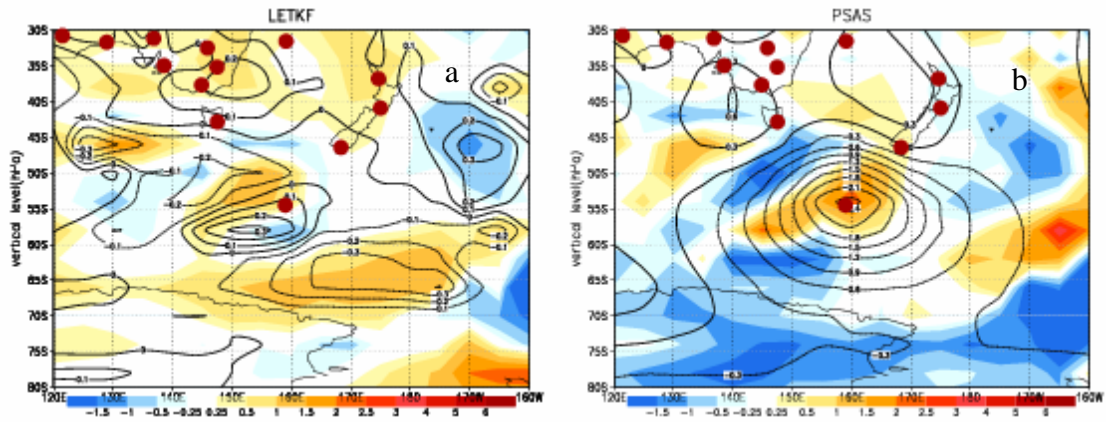

Fig. 9 500hPa temperature analysis increments (contour) and background error (shaded) for LETKF (left panel) and PSAS (right panel) at 12z Feb 12[th]. The dots represent the rawinsonde observation locations which are geopotential height observations. Both schemes assimilate winds and geopotential height observations.

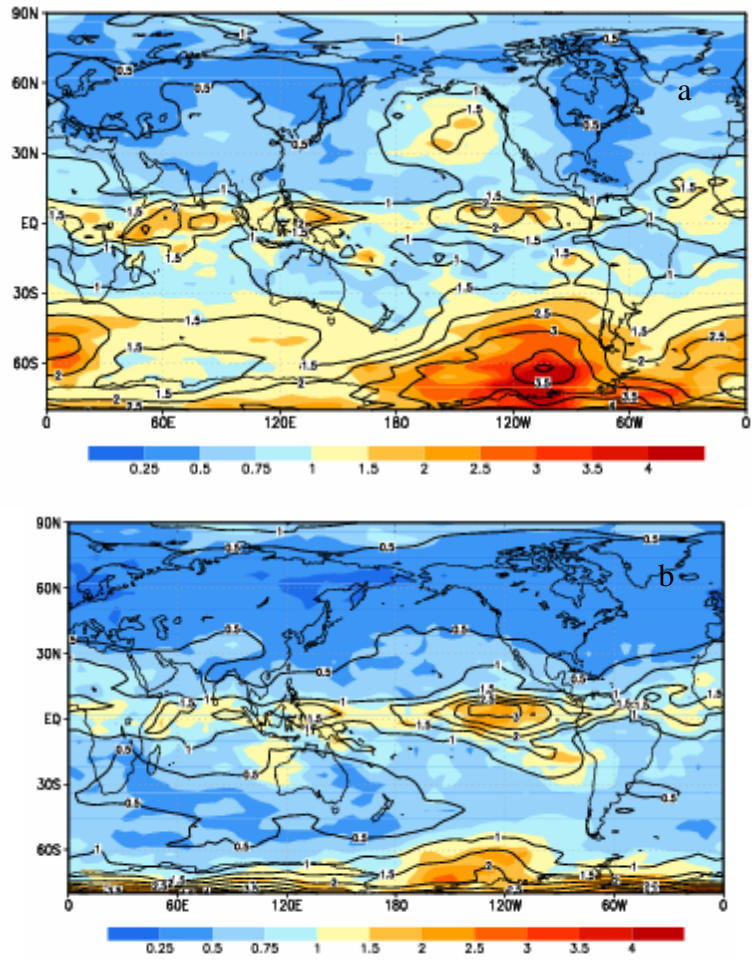

Fig.10 Average ensemble spread of zonal wind (averaged over February, contour; Unit: m/s) and the ensemble mean error (shades; Unit: m/s) at 500hPa ( (a) is calculated from LETKF with wind and geopotential height observations assimilated, (b) is calculated from LETKF with wind, temperature and surface pressure observations assimilated).

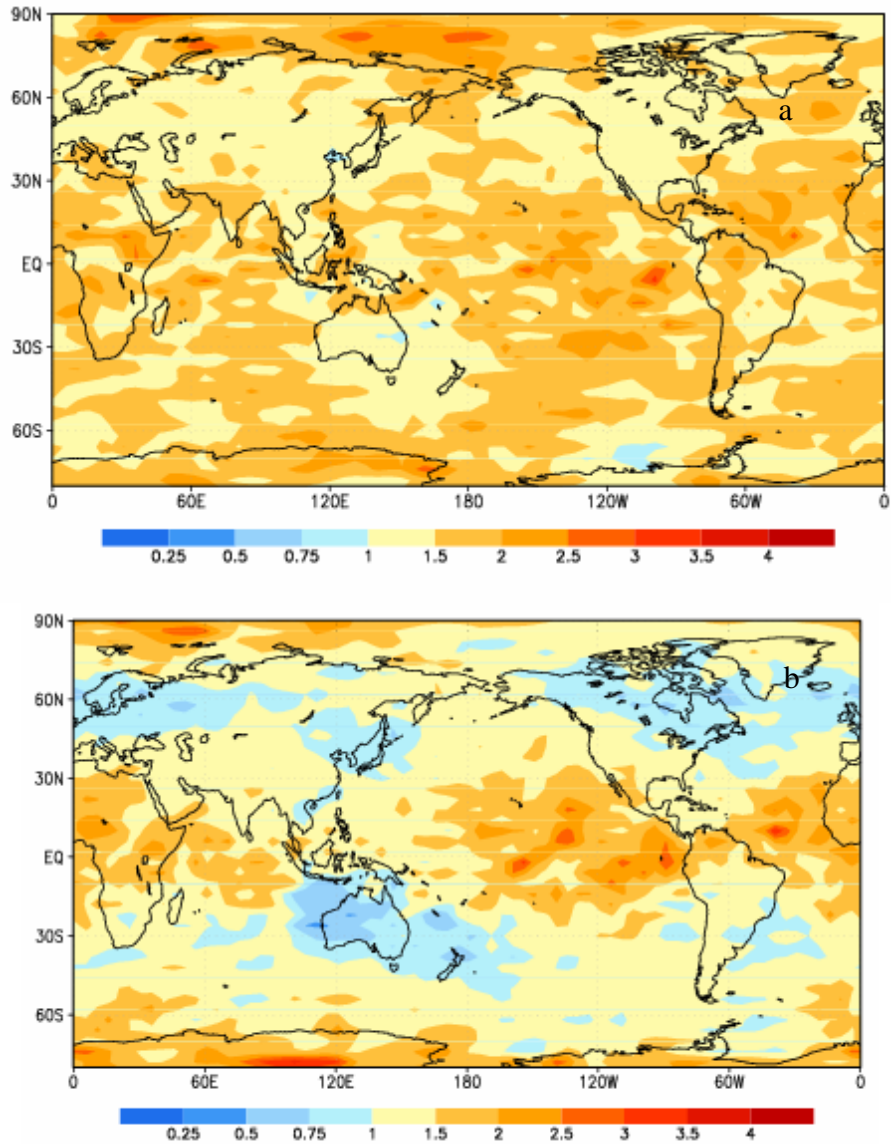

Fig. 11 The ratio of time average ensemble spread and ensemble mean error of zonal wind at 500hPa ((a) is calculated from LETKF with wind and geopotential height observation assimilated, (b) is calculated from LETKF with wind, temperature and surface pressure observations assimilated).